# Ultrashort Echo Time Magnetic Resonance Fingerprinting (UTE-MRF) for Simultaneous Quantification of Long and Ultrashort $T_2$ Tissues

Qing Li[1], Xiaozhi Cao[1], Huihui Ye[1,2], Congyu Liao[1], Hongjian He[1], and Jianhui Zhong[1,3]

[1] Center for Brain Imaging Science and Technology, Key Laboratory for Biomedical Engineering of Ministry of Education, College of Biomedical Engineering and Instrumental Science, Zhejiang University, Hangzhou, Zhejiang, China

[2] State Key Laboratory of Modern Optical Instrumentation, College of Optical Science and Engineering, Zhejiang University, Hangzhou, Zhejiang, China

[3] Department of Imaging Sciences, University of Rochester, Rochester, USA

Funding information: This work was supported by National Key R&D Program of China, Grant/Award Numbers: 2017YFC0909200, 2017YFE0104000, 2016YFC1300302; National Natural Science Foundation of China, Grant/Award Numbers: 91632109, 81871428, 61701436, 61525106, U180920013, 81873908; Shenzhen Innovation Funding, Grant/Award Numbers: JCYJ20170818164343304 and JCYJ20170816172431715; Major Scientific Project of Zhejiang Lab, Grant/Award Numbers: 2018DG0ZX01.

Corresponding author: Jianhui Zhong (email: jzhong3@gmail.com) , Room 202A, Zhouyiqing Building, Yuquan Campus, Zhejiang University, No. 38 Zheda Road, Hangzhou, China, 310027.

Abstract

**Purpose**: To demonstrate an ultrashort echo time magnetic resonance fingerprinting (UTE-MRF) method that allows quantifying relaxation times for muscle and bone in the musculoskeletal system and generating bone enhanced images that mimic CT scans.

**Methods**: A fast imaging steady-state free precession MRF (FISP-MRF) sequence with half pulse excitation and half projection readout was designed to sample fast $T_2$ decay signals. Varying echo time (TE) of a sinusoidal pattern was applied to enhance sensitivity for tissues with short and ultrashort $T_2$ values. The performance of UTE-MRF was evaluated via simulations, phantom and in vivo experiments.

**Results**: A minimal TE of 0.05 ms was achieved. Simulations indicated the sinusoidal TE sampling increased $T_2$ quantification accuracy in the cortical bone and tendon but had little impact on long $T_2$ muscle quantifications. For the rubber phantom, the averaged relaxometries from UTE-MRF ($T_1$=162 ms and $T_2$=1.07 ms) compared well with the gold standard ($T_1$=190 ms and $T_2^*$=1.03 ms). For the long $T_2$ agarose phantom, the linear regression slope between UTE-MRF and gold standard was 1.07 ($R^2$=0.991) for $T_1$ and 1.04 ($R^2$=0.994) for $T_2$. In vivo experiments showed the detection of the cortical bone (averaged $T_2$=1.0 ms) and Achilles tendon (averaged $T_2$=15 ms). Scalp structures from the bone enhanced image show high similarity with CT.

**Conclusion**: UTE-MRF with sinusoidal TEs can simultaneously quantify $T_1$, $T_2$, proton density, and $B_0$ in long, short, even ultrashort $T_2$ musculoskeletal structures. Bone enhanced images can be achieved in the brain with UTE-MRF.

# Introduction

Magnetic resonance imaging (MRI) provides exquisite contrast in soft tissues. However, bone, tendon, meniscus, and myelin sheath tissues, which have short or ultrashort $T_2$ values (on the order of 1-10 ms), are barely detectable in conventional MRI (1-3). Ultrashort echo time (UTE) and zero echo time (ZTE) imaging techniques have been used to detect these constituents (4-6). Due to their long scan time, single-point (7,8) and multipoint (9) imaging techniques are rarely used for in vivo studies. To increase k-space coverage efficiency, center-out readout trajectories, such as radial and spiral, are employed (10-12) in association with half-pulse slice selective excitation for 2D imaging or nonselective hard pulse excitation for 3D imaging.

Quantitative UTE imaging has shown extensive applications in diagnosing cartilage degeneration (13), meniscus tears (14), age-related cortical bone deterioration and osteoporosis (15). The cortical bone water concentration, a new metric for the quality of human cortical bone, requires measurements of both $T_1$ and $T_2^*$. However, separate acquisitions of $T_1$ and $T_2^*$ maps typically take more than an hour (1). To reduce the scan time, Abbasi-Rad et al. quantified $T_1$ using a dual TR method and employed an assumed $T_2^*$ value (15), with obvious omission of $T_2^*$ differences between healthy subjects and patients.

UTE and ZTE techniques are not only applied in MRI systems to diagnose musculoskeletal diseases but also in positron emission tomography (PET)/MRI systems to produce pseudo-CT (pCT) images for PET attenuation correction. Bone contrast is enhanced by suppressing background long $T_2$ signals using, e.g., a long $T_2$ suppression pulse (6,16), short $T_2$ selective double inversion recovery preparation (17), or image subtraction at two echo times (18). Wiesinger et al. obtained pCT images by segmenting the proton density (PD) map using a ZTE sequence (19). Since soft tissue, bone, and air possess different relaxometries and PDs, simultaneous characterization of multiple tissue properties would be beneficial for highlighting bone structures.

Magnetic resonance fingerprinting (MRF) (20), which simultaneously quantifies multiple tissue properties, could benefit the characterization of bone. Different tissue properties such as $T_1$, $T_2$, $T_2^*$, and PD are incorporated into the signal evolutions induced by changing the flip angle, repetition time (TR), and echo time (TE) (21). In addition, MRF has shown its potential clinical applications in the human abdomen and brain (22,23). However, challenges remain in using MRF to quantify tissues containing ultrashort $T_2$ components. First, since the original MRF method has a minimum echo time of several milliseconds, ultrashort $T_2$ / $T_2^*$ tissues are barely detected. Second, the low proton density in ultrashort $T_2$ tissue reduces the overall signal and thus degrades the accuracy of MRF dictionary matching. Third, to avoid $T_2^*$ blurring, an acquisition window of 0.81 $T_2$ is recommended for 2D imaging (24), which constrains the acquisition window length of cortical bone to submilliseconds. Fourth, since a radial spoke contains less information than a typical spiral arm, several uniformly rotated spokes are needed at each MRF time point (25).

In this work, an ultrashort echo time MR fingerprinting (UTE-MRF) method is proposed to address the problems listed above. The proposed method is based on a fast imaging with steady-state precession (FISP) sequence, which has been demonstrated in previous MRF studies (21,26). In the proposed UTE-MRF sequence, sinc excitation pulses are replaced by self-refocused half pulses (27,28). When combined with the golden angle half-spoke radial projection (29), the minimum echo time is reduced to 0.05 ms. FAs and TEs vary from TR to TR to generate fingerprint-like signal evolutions for long, short, and ultrashort $T_2$ tissues. Multiple repetitions (reps) are performed to increase the SNR for ultrashort $T_2$ tissues. Simulations, phantom studies, and in vivo studies were performed to test the proposed UTE-MRF method with respect to tissue quantification and bone enhanced image synthesis.

## Methods

**Pulse sequence design**

A diagram for the 2D UTE-MRF method is shown in Fig. 1 based on the FISP-MRF sequence (26). To reduce the echo time, a 0.7-ms half pulse was employed, which was generated by dividing a 1.2-ms sinc pulse with a time bandwidth of 6 into 2 subpulses. The VERSE algorithm (30) was applied to cool down the peak RF amplitude and the corresponding interrupted slice selection gradient. The half pulses were accompanied by the positive/negative polarized bipolar slice selection gradient to achieve complete slice selection (11,28) and reduce the impact of the eddy current (31). An MRF unit contained 480 images with variable FAs and TEs (TEmin = 0.05 ms, TEmax = 0.6 ms, and TE variation period = 120). The FA pattern consisted of four half-period sine waves whose peak FAs were 32°, 22°, 60°, and 10°, and the minimum FA was 5°. TR was fixed at 6 ms.

With ramp sampling, the readout window was 0.79 ms (0.64 ms for the plateau and 0.15 ms for the ramp), and the readout bandwidth was 1780 Hz/pixel. A waiting time of 3 s was applied after MRF unit 1 to recover the spin to its initial state before MRF unit 2. Delay time between MRF repetitions was 3 s. To maximize the SNR while constraining the scan time to less than 1 min, UTE-MRF was repeated 5 times for phantom, ankle, and brain imaging. An additional repetition was performed to increase the SNR for bone quantification, leading to a scan time of 68 s. To increase data incoherence and reduce the eddy current effect within a repetition, the radial spokes were rotated at a small golden angle (23.62°) (29,32). Among repetitions, the radial trajectories were uniformly distributed over a unit circle.

**Direct B₀ map estimation from the phase of MRF frames**

In previous MRF studies, echo time variations were applied to either increase $T_2^*$ sensitivity (33) or to separate water and fat signals (25,34). In this study, we employed a sinusoidal TE varying pattern to increase MRF sensitivity for short and ultrashort $T_2$ tissues. Fig. 2 shows the signal evolutions of six tissue constituents (under the same $T_1$ of 180 ms but increased $T_2$ values of 0.5, 1, 2, 5, and 10 ms) simulated with the extended phase graph (EPG) (35,36) and FA pattern in Fig. 1b. Intuitively, normalized signal evolutions under constant TEs (first two columns in Fig. 2) are difficult to differentiate, even when TE is minimized to 0.05 ms. Sinusoidal TE sampling (from 0.05 to 2 ms) increases the signal evolution differences. However, variable TE causes a spatially and temporally dependent phase based on field offset and TE. To prevent these phase disturbances, Rieger et al. only used the magnitude of both measured data and dictionary entries (33). To avoid loss of phase information, off-resonance-induced phase error was corrected by either employing a prescanned B₀ map (37), or incorporating the B₀ effects into the MRF dictionary (34). However, these methods add extra computation in MRF dictionary generation and matching processes. To avoid such shortcomings, we adapted a dictionary-free B₀ estimation method based on the sinusoidal TE pattern, an idea inspired by the amplitude modulation and demodulation that has been widely employed in communication systems (38). The phase accumulated in MRF image series is modulated by the field offset $B_{off}$ and sinusoidal TE carrier wave $(\alpha sin(\omega\tau) + \beta)$,

$$Phase = 2pi \cdot B_{off} \cdot (\alpha sin(\omega\tau) + \beta) + n, \qquad [1]$$

where $B_{off}$ is the frequency offset (unit: rad) caused by field inhomogeneity and the chemical shift; $\alpha$, $\beta$, and $\omega$ are TE sampling parameters, where $\alpha = (TE_{max} - TE_{min})/2$, $\beta = (TE_{max} + TE_{min})/2$, and $\omega$ is the frequency of the sinusoidal waveform; $\tau$ is the time vector $[1, 2, \cdots, F]^T$ (unit: TR); and $n$ represents the noise term. To estimate $B_{off}$, the accumulated phase in [1] is demodulated via multiplying the carrier wave $sin(\omega\tau)$,

$$dPhase = \left(2pi \cdot B_{off} \cdot (\alpha \sin(\omega\tau) + \beta) + n\right) \cdot \sin(\omega\tau). \quad [2]$$

$dPhase$ is then derived as

$$dPhase = 2pi \cdot B_{off} \cdot \alpha \cdot \left(\frac{1+\cos(2\omega\tau)}{2}\right) + \left(2pi \cdot B_{off} \cdot \beta + n\right) \cdot \sin(\omega\tau). \quad [3]$$

The noise term $n$ in [3] comprises physiological, thermal, and other MRI system-related noise. However, since phase maps are reconstructed from single k-space interleaving at each TR, the noise $n$ consists mostly of data undersampling aliasing. To reduce k-space undersampling-induced phase noise, a sliding window matrix $S$ is applied (39) and multiplied from both sides of [3],

$$S \cdot dPhase = 2pi \cdot S \cdot B_{off} \cdot \alpha \cdot \left(\frac{1-\cos(2\omega\tau)}{2}\right) + S \cdot \left(2pi \cdot B_{off} \cdot \beta + n\right) \cdot \sin(\omega\tau) \quad [4]$$

where $S$ is an $F \times F$ array containing values of either 0 or 1. Each row at $S$ behaves as a window. Only if the element is within the window, its value is 1; otherwise, 0. From the first row to the last row in $S$, the window moves from left to right. For a sliding window matrix with a window size of 4, we have $S$ in the form of

$$S = \begin{pmatrix} 1 & 1 & 1 & 1 & 0 & \cdots & 0 & 0 \\ 0 & 1 & 1 & 1 & 1 & \cdots & 0 & 0 \\ 0 & 0 & 1 & 1 & 1 & \cdots & 0 & 0 \\ & & & \vdots & & & & \\ 0 & 0 & 0 & 0 & 0 & \cdots & 1 & 1 \end{pmatrix}_{F \times F} \quad [5]$$

Due to the nature of the smoothed phase maps (shown in the supplementary video), the noise $n$ shows little impact and is ignored after the sliding window. A time integral among constant TE periods $mT$ ($T = 2\pi/\omega$, and $m$ is the number of periods) is employed as a lowpass filter, such that constant phase accumulations carried by $\cos(2\omega\tau)$ and $\sin(\omega\tau)$ are neutralized. The field offset $B_{off}$ is derived as follows:

$$B_{off} = \frac{2}{2pi \cdot \alpha \cdot mT} \cdot \sum_0^{mT}(S \cdot dPhase) \quad [6]$$

**Image reconstruction**

MRF dictionaries were generated under variable FAs in Fig. 1b and sinusoidal TEs (TEmin = 0.05 ms, TEmax = [0.05:0.05:1.0] ms) using EPG. Since our previous work have demonstrated that the TE variation period had little impact on the tissue quantification (40), the TE variation period was set to 120 TRs. Dictionary $T_1$ values varied from 10 to 3000 ms, including [10:10:400, 400:20:2000, 2000:40:3000] ms. The $T_2$ values varied from 0.1 to 300 ms, including [0.1:0.1:5, 5:5:150, 150:10:300] ms. A two-step dictionary generation method was performed to model the partial recovery effect (41,42). The dictionary entries were normalized.

Parametric maps of $T_1$, $T_2$, PD, and $B_0$ were generated in four steps. First, multichannel images were transformed from k-space into image space using inverse NUFFT (43) and combined using adaptive coil combination (44). Images from positive and negative excitations were directly complex summed. Second, the $B_0$ map was estimated from MRF phase maps based on [6], where a sliding window length of 20 was employed to balance the reduction of undersampling aliasing and tissue specificity lost. Meanwhile, phase maps from frame No. 240 to 480 (i.e., 2 TE variation periods) were used to avoid the superposition of IR-induced phase inversion. Third, to remove phase accumulations induced by field inhomogeneity, complex MRF images from step 2 were multiplied by the conjugate phase maps depending on $B_0$ and TE. Fourth, image frames were normalized and their dot-product with dictionary entries were calculated (20). Since cortical bone has low spin density and ultrashort $T_2$, it is sensitive to bone marrow signals from radial undersampling artifacts. Here, we first employed the partial volume dictionary matching method (45) to estimate and remove bone marrow components and then applied MRF dictionary matching to obtain the cortical bone $T_2$.

Since an adiabatic IR pulse was utilized, the MRF possessed an intrinsic advantage in long $T_2$ suppression near the soft tissue IR-null point. Although the ultrashort $T_2$ relaxation time of bone makes it decay quickly in the transverse plane, it recovers rapidly and shows longitudinal hyperintensity due to its short $T_1$. With the calculated $T_1$ and $T_2$ maps, transverse and longitudinal magnetization maps at any MRF frame can be generated by reference to the dictionary. The skull enhanced images were obtained by selecting a proper frame in the longitudinal magnetizations where ultrashort $T_2$ bone was highlighted.

**Simulation study**

To investigate the impact of the sinusoidal TE sampling on the accuracy of tissue quantification, numerical simulations were carried out on a digital phantom (size of 110 x 110) shown in Fig. 3, including muscle, tendon, total water of bone, and free water of bone. Dark space in the phantom was for air. Most of the tissue properties (listed in Fig. 3b) were chosen following the literature (17,46,47). Ideal MRF frames were generated from MRF transverse magnetizations using EPG. Although the TE variation increased the specificity of the MRF to short $T_2$, it introduced a TE-dependent $T_2^*$ weighting. Since the impact of $T_2$ was modeled into MRF signal evolutions through EPG, the $T_2^*$-caused decay, $\exp(TE/T_2^*) = \exp(TE/T_2+TE/T_2')$, was incorporated by multiplying the ideal MRF frames by an additional exponential $T_2'$ decay. The nominal $T_2'$ values, which are shown in Fig. 3b, led to a $T_2^*$ of 25 ms for muscle, 2.3 ms for tendon, 0.7 ms for the total water of bone, and 2.4 ms for the free water of bone.

SNR variations were also considered. Complex white Gaussian noise was added to represent different SNR levels as listed in Fig. 3b. k-Space samples were obtained point-by-point through forward NUFFT of the noised MRF frames, the golden angle rotated radial readout and the corresponding density compensation function (48). TEmin was fixed at 0.05 ms, and TEmax was increased from 0.05 to 1 ms in steps of 0.05 ms. Multiparametric maps were reconstructed following the method in the image reconstruction section. The root mean square error (RMSE) of the measured $T_1$ and $T_2$ maps was calculated compared to the gold standard $T_1$ and $T_2$ values.

**Phantom and in vivo experiments**

Experiments were conducted on a whole-body 3T scanner (Magnetom Prisma, Siemens Healthineers, Erlangen, Germany). A homemade phantom (Fig 5) with agarose tubes (mimicking soft tissues) and rubber plugs (mimicking ultrashort $T_2$ tissues) was imaged using a 20-ch head coil. Seven of those tubes were filled with different concentrations of $MnCl_2$ to represent tissue components of variable $T_1$ and $T_2$ values, and one tube was filled with vegetable oil (94% soybean oil and 6% sunflower oil), which has one dominant resonance peak at about 3.46 ppm from water (see Supporting Information Figure S1) similar as Reeder et al. reported (49). In vivo experiments were carried out with the approval of the local institutional review broad. A healthy volunteer participated in calf and tendon imaging using a 15-ch knee coil. Brain scans were performed on 3 healthy volunteers and a patient with facial neuromas using a 64-ch brain coil. For the healthy volunteers, the FA trains in Fig. 1a were used. The SAR reported by the scanner was 48% of the safety threshold under Normal mode. For the patient scan, the FAs were halved for safety considerations. The bone enhanced image from the patient was compared with CT images (tube voltage = 80 kV, tube current = 365 mA, DLP = 97.9 mGy·cm, mean dose = 0.34 mSv, slice thickness = 1.0 mm, and resolution = 0.5 x 0.5 $mm^2$) acquired and reconstructed on a Philips iCT scanner (Brilliance iCT, Philips Healthcare, The Netherlands) for her surgery planning at a local hospital. The 3D CT images were rotated to match the orientation of the MRI image using the function of 3D volume rendering in RadiAnt Dicom Viewer (Medixant Co., Poland).

The slice thickness was 6 mm for the phantom and 7 mm for the in vivo study. To obtain a high-quality bone enhanced image, TE was minimized to 0.05 ms and was unchanged for the brain scan. Multiparametric maps were reconstructed to achieve a resolution of 1.0 x 1.0 $mm^2$ (matrix size of 240 x 240) for the phantom, leg, and tendon and 0.75 x 0.75 $mm^2$ (matrix size of 256 x 256) for the brain using MATLAB R2014a (The MathWorks, MA) on a Linux (Red Hat Enterprise) server (with Core i7 Intel Xeon 2.8 GHz CPUs and 64 GB RAM).

An inversion recovery UTE (IR-UTE) sequence was used to acquire a gold standard $T_1$ map with inversion times (TIs) = 50, 100, 200, 400, and 800 ms, TR = 3000 ms, TE = 0.05 ms, and number of spokes = 248. The gold standard $T_2^*$ map was generated from a UTE sequence with TEs = 0.05, 0.2, 0.5, 1, 2, and 4 ms, TR = 1500 ms, and number of spokes = 248. All images acquired with the IR-UTE and UTE sequences were reconstructed following the first step of MRF image reconstructions. The gold standard $T_2$ map was measured with a spin echo (SE) sequence with TEs = 25, 50, 75, 100, and 125 ms, TR = 3000 ms, matrix size = 192 x 192, resolution = 1 x 1 mm$^2$, and 6/8 partial Fourier acquisition. The total acquisition time was 124 min for $T_1$, 74 min for $T_2^*$, and 36 min for $T_2$. $T_1$, $T_2$, and $T_2^*$ maps were reconstructed using a toolbox (http://www-mrsrl.stanford.edu/~jbarral/ t1map.html) (50).

## Results

**Simulation**

Fig. 4 shows the simulation results of the musculoskeletal tissue mimic phantom under 3 and 6 MRF reps. Overall, the $T_1$ and $T_2$ RMSEs are decreased by approximately twofold with twice repetitions. The $T_1$ and $T_2$ quantification of muscle are robust against TE variations. Tendon and bone total water show higher improvement in $T_2$ quantification with extended TE sampling in 3 reps than in 6 reps. $T_1$ RMSEs are slightly increased from 33/17/77 ms to 36/21/82 ms for tendon/total water/free water under 6 reps. Free water shows the largest quantification errors due to its lowest PD. When $T_1$ and $T_2$ maps reconstructed at TEmax = 0.6 ms are compared with the gold standard $T_1$ and $T_2$ (Fig. 4b), strong streaking artifacts are observed with 3 reps, especially in the $T_2$ difference map. When the repetitions are increased to 6, $T_2$ quantification errors are substantially reduced, but $T_2$ biases are observed in tendon and free water. The average measurements from UTE-MRF under 6 reps with TEmax = 0.6ms is compared with nominal relaxometries in Supporting Information Table S1. UTE-MRF shows high-degree agreements with nominal $T_1$. However, the $T_2$ values from UTE-MRF tend towards the nominal $T_2^*$ from the longest $T_2$ in muscle to the shortest $T_2$ in bone total water.

**Phantom and In Vivo Studies**

Fig. 5 shows the phantom quantitative maps from the UTE-MRF and gold standard methods. Since rubber is undetectable in the SE sequence, the gold standard $T_2^*$ is measured instead of $T_2$ for rubber and compared with the quantified $T_2$ from UTE-MRF. The $T_1$, PD, and $T_2$ (agarose phantom) maps agree well with the gold standard. Meanwhile, the rubber $T_2$ measured with UTE-MRF is close to the gold standard $T_2^*$. The oil $T_2$ is approximately 120 ms accompanied with a ~400 Hz frequency shift detected by UTE-MRF, which is close to the reported value of fat (44), but is underestimated in SE method.

Fig. 6 shows the ROI-based comparisons between UTE-MRF and the gold standard. The $T_2$ from UTE-MRF is compared with the gold standard $T_2^*$ for the ultrashort $T_2$ rubber phantoms (red markers) and with the gold standard $T_2$ for the other long $T_2$ agarose tubes (green markers). For the long $T_2$ agarose phantoms the $T_1$ and $T_2$ values from UTE-MRF show good agreement with the gold standard $T_1$ ($y = 1.07x - 43.71$, $R^2=0.991$) and $T_2$ ($y = 1.04x - 3.27$, $R^2=0.994$). Fig. 6b lists the $T_1$ and $T_2$ values of UTE-MRF from the chosen ROIs.

Fig. 7 shows in vivo ankle and leg results from UTE-MRF. Since $T_2$ varies from several microseconds in bone to one hundred or more microseconds in bone marrow, the $T_2$ maps are displayed separately on two scales for long and short $T_2$ tissues. Consequently, the long $T_2$ tissues, which are clearly observable in the third column, seem to be saturated at the finer display scales in the fourth column. Two ROIs (5 x 5) in bone marrow and muscle are marked by red boxes in the anatomic image. The mean $T_1$ / $T_2$ for bone marrow and muscle are 364/141 ms and 1038/27 ms, respectively. Short $T_2$ tissues including the tibia (labeled by *a*), fibula (*b*), Achilles tendon (*c*), and peroneus longus tendon (*d*) are observed in the fourth column under different display ranges. The average $T_2$ in a 3 x 3 ROI in Fig. 7 is 1.0 ms for cortical bone and 15 ms for Achilles tendon. High off-resonance artifacts are detected in bone marrow and skin due to the chemical shift between fat and water.

Fig. 8 shows the simulation results for the brain for the longitudinal magnetization time evolution using $T_1$ and $T_2$ relaxometries reconstructed from UTE-MRF. As shown in the figure, since tissue relaxometries are quantified through dictionary matching, one can look up the dictionary and generate transverse / longitudinal magnetization changes at any MRF time point.

Fig. 9 shows brain bone enhanced images and parametric maps from 3 healthy volunteers using UTE-MRF. Although UTE-MRF loses its sensitivity to ultrashort $T_2$ quantification when TE is fixed at 50 µs, it gains SNR and achieves higher resolution. The bone enhanced images show enhanced skulls, which were generated by dividing the Mz map reconstructed from MRF frame No. 130 (Fig. 8) by the $T_1$ map. Some dark spots are detected in both the $T_1$ and $T_2$ maps, especially at the lateral ventricle. Those spots (as indicated by the blue arrows) show high intensity in the bone enhanced images in S3. The results from three orthogonal planes are shown in Supporting Information Figure S2.

Bone enhanced images from the patient with facial neuromas are compared with CT images in Fig. 10. Streaking artifacts are observed in the $T_2$ maps, especially at the middle of the brain. The zygomatic bone and hyperintensity areas in the bone enhanced images are in good agreement with the CT images, as indicated by the red arrows.

**Discussion**

In this work, a UTE-MRF method was proposed to simultaneously quantify long and short / ultrashort $T_2$ tissues and to synthesize bone enhanced images. The sinusoidal TE pattern was incorporated with a minimal TE of 50 µs. Variable TE-induced phase changes were compensated using the $B_0$ map demodulated from the phases of the MRF frames. $T_1$, $T_2$, and PD maps were reconstructed through dictionary matching. The longitudinal magnetization map, which was obtained by looking up the dictionary based on the quantified relaxometries, was employed to produce bone enhanced images. Simulation, phantom measurements, and in vivo studies showed the capability of UTE-MRF for detecting and quantifying musculoskeletal system mimic phantoms, tendon, cortical bone, and muscle. In vivo brain scans also demonstrated its capability for brain quantifications and producing bone enhanced images.

With the sinusoidal TE variation, the $B_0$ map can be directly estimated from the phase of MRF frames through the amplitude demodulation method without an additional dictionary and matching burden. TE variation causes a synchronized sinusoidal phase change as the product of TE and $B_0$ offset caused by chemical shift and field inhomogeneity. Calculating the field changes directly using [6] suffers from strong undersampling aliasing, resulting in poor field estimation. Here, a noniterative sliding window method (39) was used to reduce undersampling artifacts. Cao et al. chose the number of fully sampled spiral interleaves as the window length. For imaging FOV = 140 x 140 $mm^2$ and resolution = 1.0 x 1.0 $mm^2$, a fully sampled radial would require 220 spokes. However, a window length of 220 would tremendously decrease the tissue specificity, as it is approximately twice the TE and FA variation period (120). To balance tissue specificity and undersampling aliasing, a window size of 20 was used in this study, and the maximum TE variation within the sliding window was (TEmax−TEmin)/2. Since multiple repetitions were performed to increase SNR for bone, tendon, and the rubber phantom, the actual image undersampling factor within the sliding window was 2.2. There is a tradeoff between $B_0$ sensitivity and image SNR in choosing TE sampling ranges. More accurate field estimation would be achieved through strong phase changes via increased TE variation. However, there is more loss of image SNR as a penalty due to the $T_2/T_2^*$ decay effect. To avoid the impact of IR-induced phase changes along MRF frames, the last 2 periods of MRF frames were applied to estimate $B_0$. Two periods (1440 ms, 2 x 120 x TR) are adequate for the tissue with the longest $T_1$ to pass through its IR null point, e.g., for a $T_1$ of 2000 ms, the IR null point is 1386 ms (ln(2) x $T_1$).

To achieve ultrashort TE, half-pulse excitation (duration of 0.7 ms) and radial acquisition (window length of 0.79 ms) were employed. Although $T_2^*$ decay during excitation has an impact on the excitation profile, Pauly has shown that there is a small loss of signal and slight broadening of the slice profile for tissue with $T_2$ of 0.25 ms and a half pulse with a duration of 1 ms (28). Since the responses of the inversion and excitation are influenced by the relaxation times, especially $T_2$ (51,52), Bloch equation with relaxation times would be more accurate to model the inversion and excitation imperfections, and these effects could be modeled into dictionary using Bloch equation simulation to further improve the quantification accuracy. Excitation performance would also be degraded by hardware imperfections. To minimize timing error between RF and gradient, the RF pulse was adjusted to perform a -10 to 10 μs (in steps of 1 μs) time shift relative to the gradient. The optimal shift time of -3 μs was chosen based on the performance of the slice profile. To decrease the impact of eddy currents, a bipolar slice selection gradient was employed based on the fact that the long-term eddy currents from ramp-up and ramp-down gradients are cancelled due to their opposite polarities (31). However, residual short-term eddy currents may still distort the slice selection gradient. Better slice selection performance could be achieved by measuring the trajectory of the slice selection gradient (11). To reduce the blurriness of the $T_2^*$ of bone ($T_2 \leq 1$ ms), a readout window of 0.79 ms was employed under an optimal window size of $0.81T_2$ (24). Off-resonance blurring was decreased by using a high readout bandwidth of 1780 Hz/pixel.

The simulation results in Fig. 4 indicate that UTE-MRF is capable of simultaneously quantifying long and short $T_2$ tissue components. $T_1$ and $T_2$ quantification errors in muscle almost remain constant with increased TE samplings, which suggests TE variations have little impact on long $T_2$ tissue. Since doubling sequence repetitions reduces $T_1$ / $T_2$ RMSEs by approximately twofold (Fig. 4a), a radial undersampling artifact is probably the primary source for the steady quantification errors. Since TE varies from frame to frame, it introduces $T_2^*$ weighting in the MRF image series.

According to the results in supplementary Table 1, the $T_2^*$ effect has little impact on long $T_2$ muscle because the maximum TE variation of 0.6 ms is much smaller than the nominal $T_2$ (32 ms) and $T_2^*$ (25 ms) values of muscle. The impact of the $T_2^*$ effect increases for shorter nominal $T_2$ and $T_2^*$ values. For the bone total water with the shortest $T_2$ and $T_2^*$, the measured $T_2$ from UTE-MRF is close to its nominal $T_2^*$ value. Although cortical bone free water has longer $T_2$ values than bone total water and tendon, its quantification accuracy is the poorest due to its low proton density. At low image SNR under 3 UTE-MRF reps, an increment of the TE sampling range could enhance the MRF specificities for ultrashort $T_2$ tissues and improve ultrashort $T_2$ tissue quantification. As suggested by the simulation performance, an optimal TEmax of 0.6 ms was used for the phantom and in vivo experiments, and the TEmin of 0.05 ms was unchanged throughout the experiments.

For the phantom quantifications in Fig. 6, most of the relaxometries in UTE-MRF were in good agreement with the gold standards, except for the $T_2$ of the No.4 rubber plug and the oil tube. In UTE-MRF, the chemical shift-caused frequency offset is encoded into the phase of the MRF frames, then estimated using [6] and compensated in the MRF frames before dictionary matching. The oil spectrum shows 6 peaks (Supporting Information Figure S1), which is similar to the Fig. 8 in the work of Reeder et al (49). Mono-exponential fitting in SE method underestimates the $T_2$ of the oil sample which actually contains multi-components. While multicomponent analysis may be needed for quantifying $T_2$, the multicomponent effect has minimal impact on UTE-MRF since the maximum TE used in UTE-MRF was 0.6 ms.

The average $T_2$ of cortical bone and tendon from the chosen ROI was 1.0 ms and 15.0 ms, respectively. Bicomponent analyses have shown that cortical bone possesses bound water ($T_2^*$ of 0.36 ms) and free water ($T_2^*$ of 5.56 ms) (53). Tendon shows similar bicomponent characteristics, with a shorter $T_2^*$ of 0.88 ms and a longer $T_2^*$ of 25.58 ms (54). The TE variations from 0.05 ms to 0.6 ms enable UTE-MRF to detect signals from both bound and free water in cortical bone and tendon, as demonstrated in the simulation results (Fig. 4b). Therefore, the $T_1$ and $T_2$ quantified from UTE-MRF are the apparent $T_1$ and $T_2$ from both short $T_2$ and long $T_2$ water components of bone and tendon.

The in vivo brain relaxometry quantification and bone enhanced images are achieved using UTE-MRF (Fig. 9 and Fig. 10). Although the sensitivity of MRF to $T_2$ was decreased by the halved FA trains in the patient study, the bone structures from UTE-MRF were still mostly consistent with the CT images. The dark spots in the in vivo $T_1$ and $T_2$ maps only appeared at areas filled with cerebrospinal fluid. Those dark spots were likely caused by inconsistent excitation due to flow effect, since two half pulses may excite different moving spins. Consequently, the mismatched flow spins may lead to signal error. To obtain high-quality bone enhanced images, TE was minimized to 50 μs in brain scans. However, the $T_2$ sensitivity of UTE-MRF for short $T_2$ tissues was reduced as a penalty. For voxels with mixed short $T_2$ and long $T_2$ components, UTE-MRF measures an apparent $T_1$ and $T_2$ from the overall signal. To resolve the MRF signal into long and short $T_2$ components, a multicomponent analysis method could be employed and modeled into dictionary (3).

The proposed UTE-MRF method was validated via simulations, phantoms, and in vivo experiments. However, there are some limitations of the current 2D UTE-MRF approach. First, this approach does not totally encompass imaging imperfections such as $B_1$ inhomogeneity and a nonideal slice profile, which limits the quantification performance. To correct $B_1$ inhomogeneity, $B_1$ maps can be acquired either independently (42) or in association with MRF data (55). To avoid slice profile imperfections, spiral acquisition-based 3D MRF has been demonstrated (41,42). Furthermore, the increase in SNR due to 3D volume excitation could also benefit the characterization of low spin density tissues such as bone. The current method still requires an acquisition time of ~1 min for a single slice, and thus the scan efficiency of UTE-MRF limits its clinical applications. The excitation time would be halved with hard pulse volume excitation in 3D (56,57). To further speed up the acquisition, effective 3D k-space sampling could be used (58). As part of the future implementation of UTE-MRF techniques, we will adapt the above and other methods to accelerate the process. Second, although the $T_2$ of cortical bone from the chosen ROI is in agreement with previous work (17), the cortical bone quantification in Fig. 7 suffers from the strong streaking artifact from bone marrow. This issue could be addressed by either separating water and fat before dictionary matching (25,37) or characterizing fat in the MRF dictionary (34). Additionally, the experiment could be optimized by optimizing UTE-MRF FA and TE variation patterns (59). Third, ultrashort $T_2$ components from the myelin membrane or membrane structures in white matter and gray matter are explored by previous UTE methods (60,61). With a TE of 0.05 ms, UTE-MRF acquires those ultrashort $T_2$ components; however, this phenomenon was not fully studied in this work. Fourth, to accomplish attenuation correction in PET/MR, the bone enhanced images must have a quantitative CT unit by assigning the CT values to different tissues segmented from MR (19,62) or a machine learning-based method (63).

## Conclusion

An ultrashort echo time MR fingerprinting (UTE-MRF) method with sinusoidal echo time variations was proposed and tested via simulations, phantom, and in vivo experiments. The capability of the method to quantify tendon and cortical bone could benefit studies of age-related bone degradation and other musculoskeletal abnormalities. Furthermore, UTE-MRF could be used in PET/MRI systems to simultaneously quantify brain tissues and aid PET attenuation correction with simultaneously synthesized bone enhanced images.

## Acknowledgments

The authors would like to acknowledge Dr. Kang Wang and Dr. Jin Jin from the First Affiliated Hospital of Zhejiang University for the help and discussions on the patient scans.

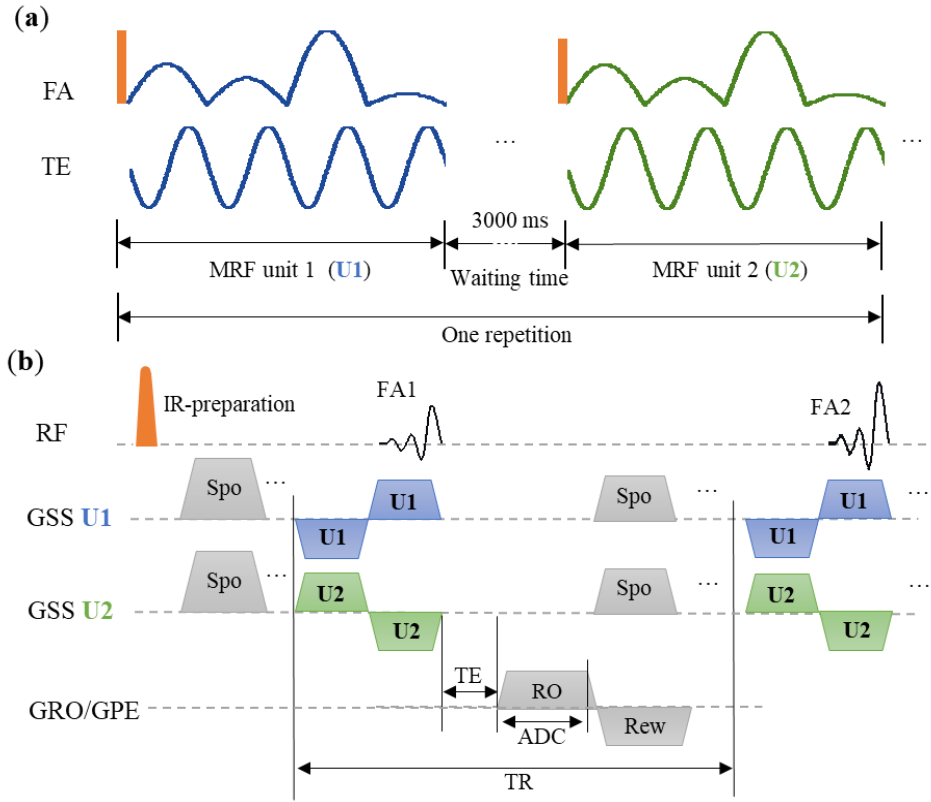

Fig. 1. Sequence diagram of UTE-MRF. UTE-MRF is acquired with the IR-FISP sequence utilizing half-pulse excitation and radial acquisition. The fingerprint-like scan shown in (a) is called an MRF unit, where the flip angle and echo time varies from TR to TR. A fixed repetition time of 6 ms was used to increase scan efficiency. One UTE-MRF repetition comprises two MRF units. Details of the RF and gradient timing in a single TR are shown in (b). The slice selection gradient polarization is fixed within one MRF unit (U1, in blue) but is inverted for the next MRF unit (U2, in green). The k-space trajectories of the same MRF frame are uniformly distributed over a unit circle when multiple repetitions are performed. Spo, RO, Rew indicate the gradients of spoiler, readout, and rewinder.

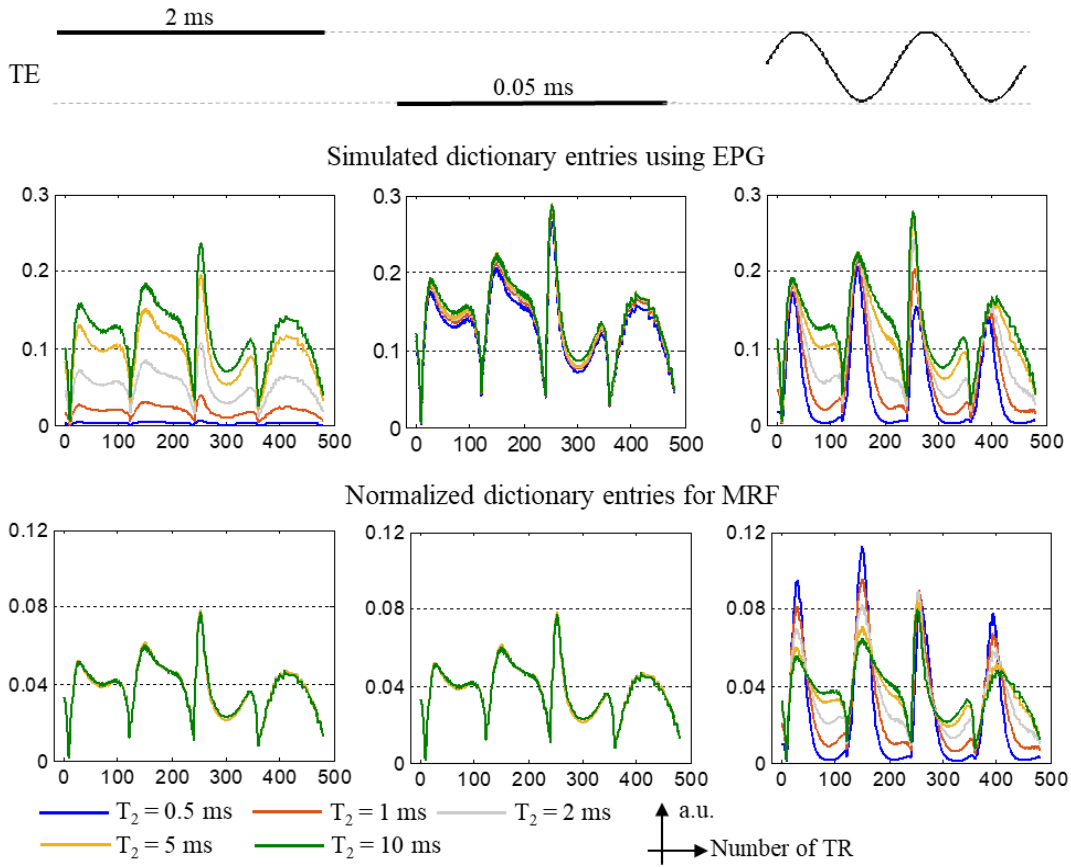

Fig. 2. Simulations of the impact of TE patterns on MRF signal evolutions (fingerprints) under the same $T_1$ of 180 ms and increased $T_2$ of 0.5, 1, 2, 5, and 10 ms. Two single TEs of 2 and 0.05 ms were used in the first and second columns, respectively. In the last column, TE was varied in a sinusoidal pattern from 0.05 to 2 ms. For dictionary entries at constant TEs (the first two columns), the difference among simulated fingerprints (second row) disappeared after dictionary normalization (last row). However, these fingerprints under sinusoidal TE could be differentiated both before and after dictionary normalization (last column).

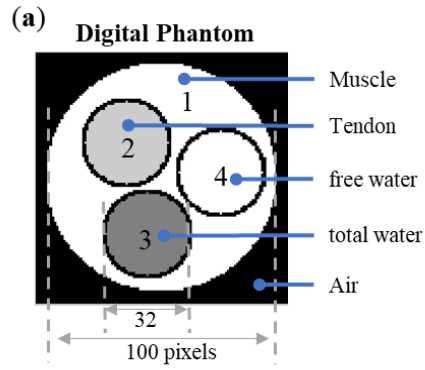

(a) Digital Phantom

(b) Tissue Parameters

|  | $T_1$ (ms) | $T_2$ (ms) | $T_2'$ (ms) | PD | SNR |
|---|---|---|---|---|---|
| Muscle | 1400 | 32 | 114 | 1 | 28 |
| Tendon | 621 | *3* | 9.9 | *0.8* | 20 |
| Total water | 246 | *1.2* | 1.7 | *0.7* | 14 |
| Free water | 524 | 3.5 | 7.6 | 0.3 | 7 |

Fig. 3. Digital phantom and its corresponding simulation parameters. The peripheral cylinder has a diameter of 100 (pixels), and the inserted small cylinders have a diameter of 32 (pixels). Tissue parameters are chosen based on the literature (16,46,47). $T_2^*$ values are 25 ms for muscle, 2.3 ms for tendon, 0.7 ms for total water, and 2.4 ms for free water

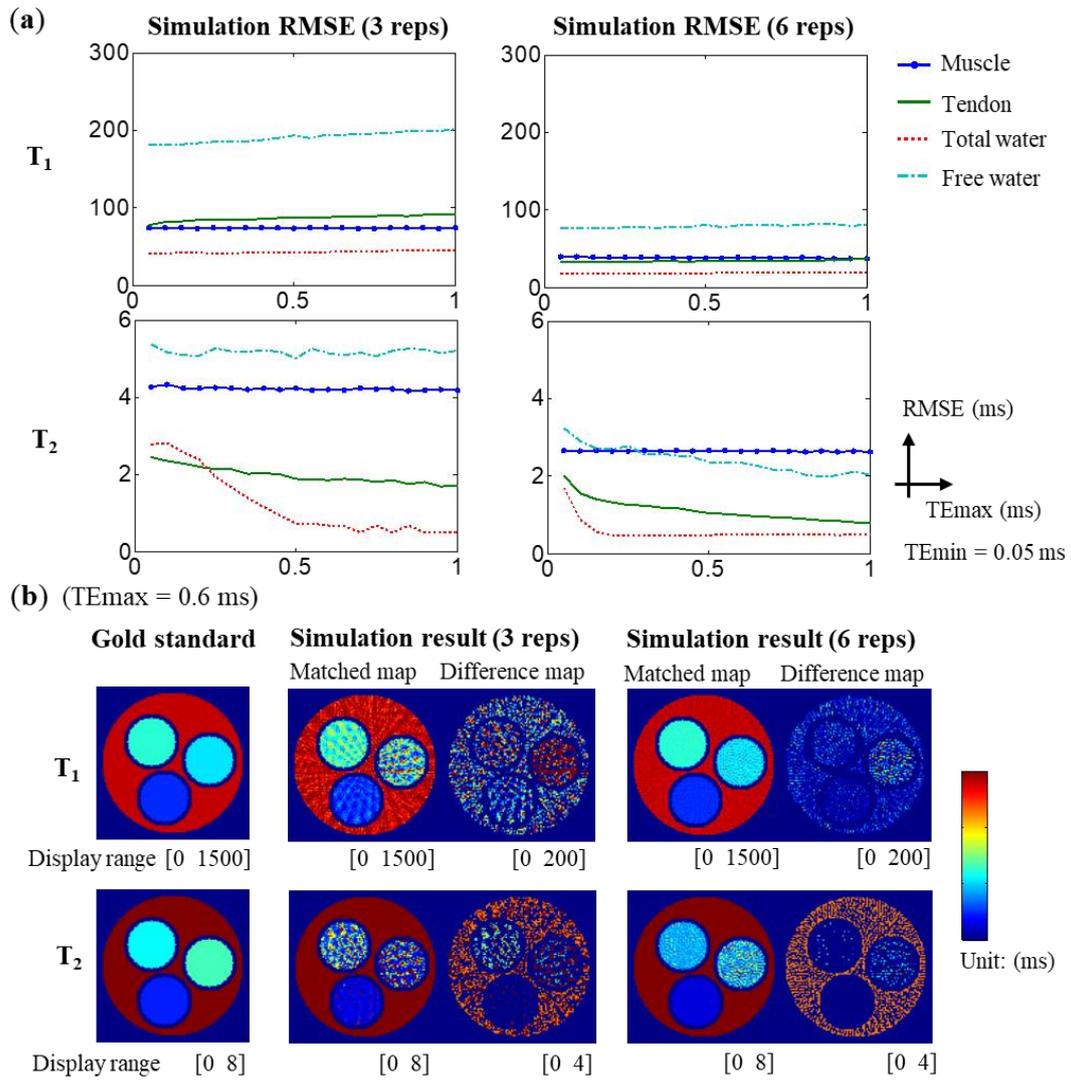

Fig. 4. Simulation results for the digital phantom with increased TE sampling ranges depending on maximum TE (TEmax), where TEmin = 0.05 ms and TEmax increases from 0.05 to 1 ms in steps of 0.05 ms. (a) shows the comparisons of the average quantification errors for each tissue under 3 and 6 repetitions (reps). As the TE sampling range increases, decreases in $T_2$ quantification errors for tendon, total water, and free water are observed. Meanwhile, the $T_1$ and $T_2$ quantification performances for long $T_2$ muscle are almost unchanged. (b) shows an example of reconstructed $T_1$ and $T_2$ maps under TEmax of 0.6 ms and their comparisons with gold standards. The streaking artifacts from the undersampled radial trajectory in 3 reps are largely reduced with 3 more reps.

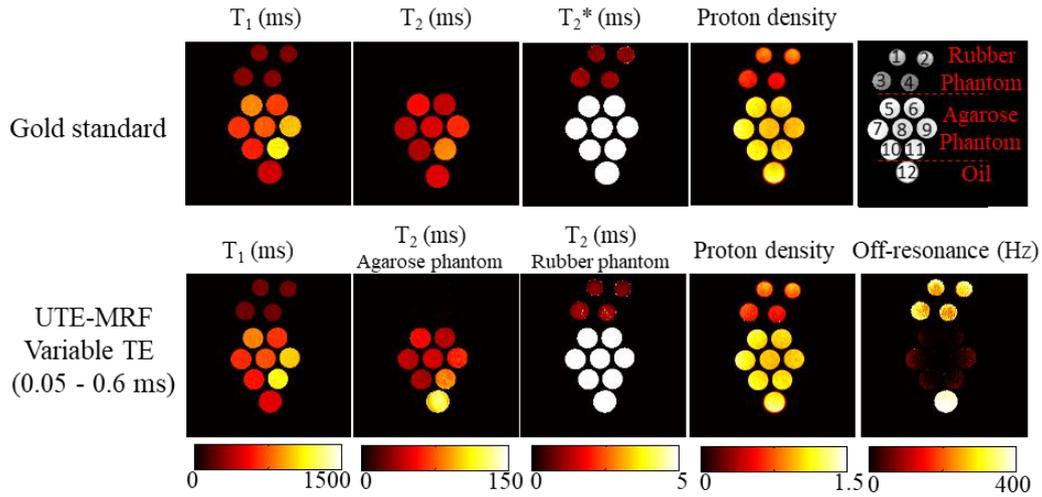

Fig. 5. Quantitative multiparametric maps in the phantom measured using the UTE-MRF (TEmin = 0.05 ms and TEmax = 0.6 ms) and gold standard methods. Each column shares the same color bar at the bottom. Samples of the phantom are indexed from 1 to 12 as shown at the upper right corner. Because the phantom has a large $T_2$ variation from approximately 1 ms to nearly 100 ms, the $T_2$ map from UTE-MRF is displayed at two scales: [0, 150] ms in the second column to visualize long $T_2$ tissues (agarose phantom and oil tube) and [0, 5] ms in the third column to visualize the short $T_2$ rubber phantom. Since the rubber signal is difficult to detect with the SE sequence under the minimum TE of 15 ms, the gold standard $T_2^*$ of the rubber phantom is used as an alternative in comparison with the $T_2$ measured from UTE-MRF. Proton density maps are normalized by the average intensity of the oil tube.

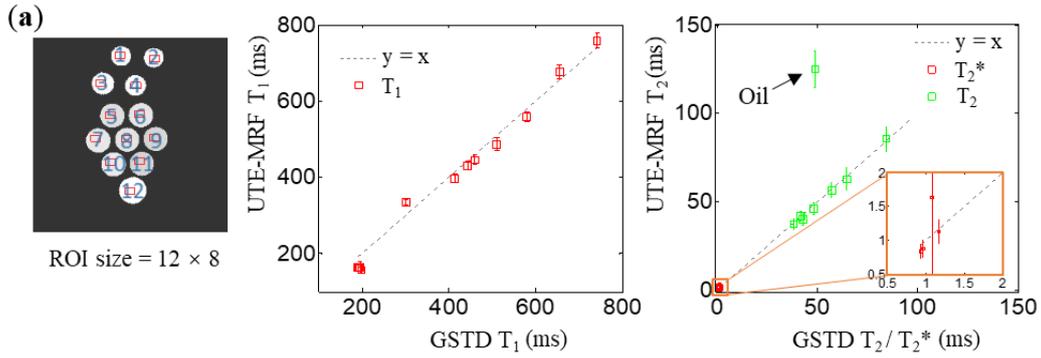

ROI size = 12 × 8

UTE-MRF Quantification Results from Chosen ROIs

| (ms) | # 1 | # 2 | # 3 | # 4 | # 5 | # 6 |
|---|---|---|---|---|---|---|
| $T_1$ | 163±10 | 164±8 | 158±9 | 164±15 | 559±13 | 446±13 |
| $T_2$ | 0.88±0.14 | 0.84±0.11 | 1.13±0.18 | 1.52±1.36 | 56±5 | 40±3 |
| (ms) | # 7 | # 8 | # 9 | # 10 | # 11 | # 12 (oil) |
| $T_1$ | 430±11 | 487±16 | 676±17 | 397±12 | 759±19 | 335±7 |
| $T_2$ | 42±3 | 45±3 | 62±2 | 37±3 | 85±7 | 122±11 |

Fig. 6. Comparisons of $T_1$ and $T_2$ quantifications between UTE-MRF (TEmin = 0.05 ms and TEmax = 0.6 ms) and gold standard from the chosen ROIs in the phantom. (a) Plots of $T_1$ and $T_2$ from UTE-MRF versus the gold standard in twelve chosen ROIs (red boxes, size of 12 x 8 pixels), where the $T_2$ values of the rubber phantom are compared with the gold standard $T_2^*$ (red marker) and the $T_2$ values of the agarose phantom are compared with gold standard $T_2$ (green marker). (b) $T_1$ and $T_2$ in UTE-MRF (mean±SD) from the chosen ROIs.

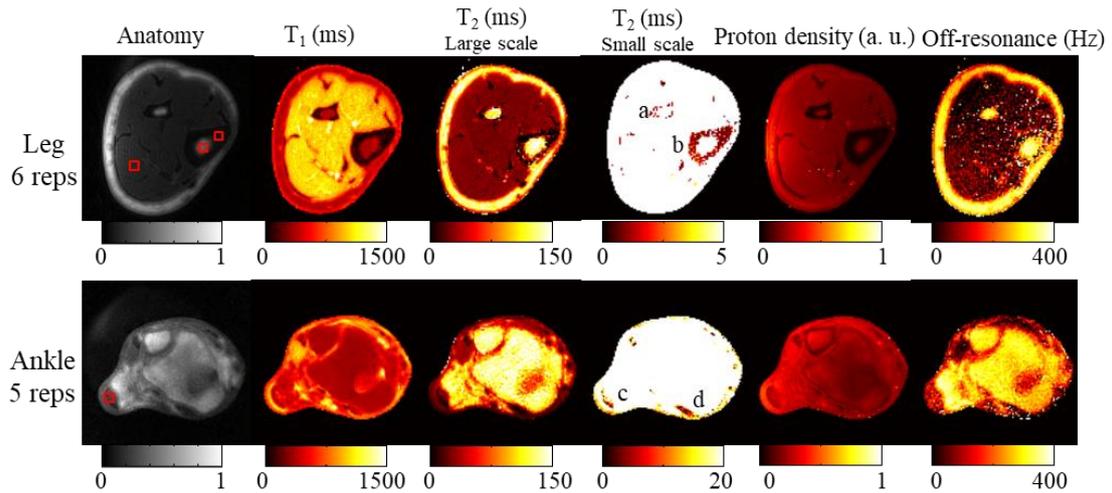

Fig. 7. In vivo results of leg and ankle (resolution = 1.0 x 1.0 mm$^2$). The images were acquired with TEmin = 0.05 ms, TEmax = 0.6 ms, TE variation period = 120, and reps = 6 for leg and 5 for ankle. Tibia (a), fibula (b), Achilles tendon (c), and peroneus longus tendon (d) are detected and quantified in the fourth column. Two 5 x 5 ROIs are selected among muscle and bone marrow and 3 x 3 ROIs among fibula and Achilles tendon. The average T$_2$ value is 1.0 ms in the chosen fibula and 15 ms in the Achilles tendon.

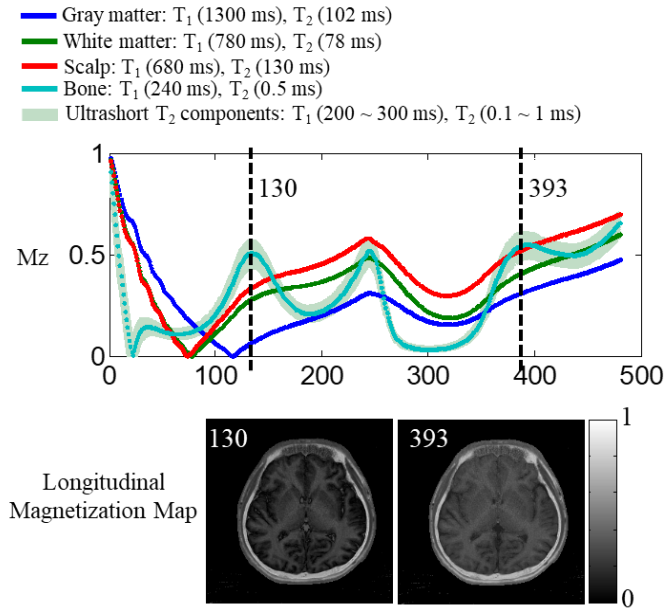

Fig. 8. Simulations of longitudinal magnetization for bone and soft tissue in the brain using EPG. Longitudinal magnetization maps at MRF No. 130 and 393 are displayed at the bottom, where bone is highlighted.

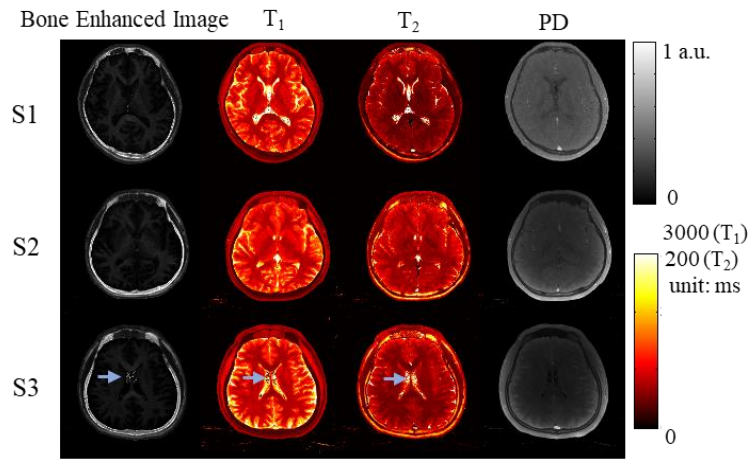

Fig. 9. Brain quantifications from 3 healthy volunteers using UTE-MRF. Volunteers are numbered from S1 to S3. Bone enhanced images and PD maps are normalized and share the same gray color bar at the upper right corner. $T_1$ and $T_2$ maps share the same color bar but with different display scales at the lower right corner.

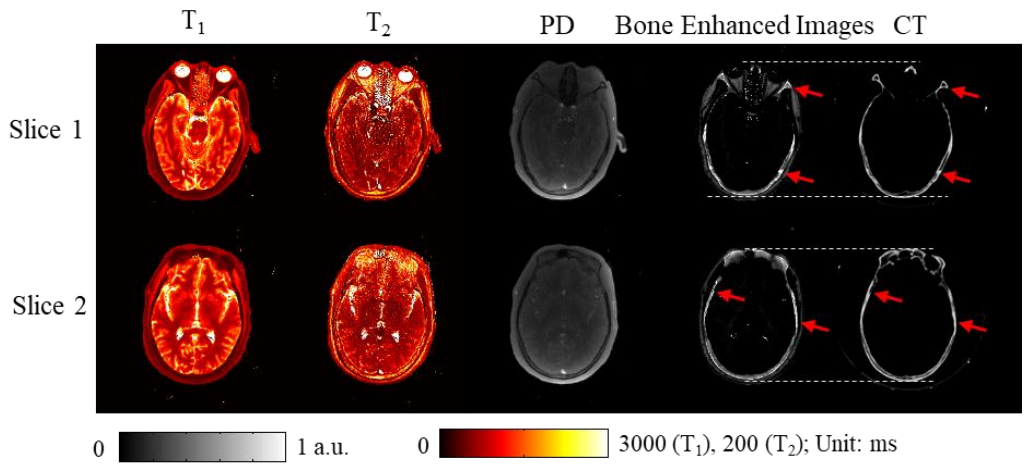

Fig. 10. Bone enhanced images, $T_1$, $T_2$, and PD maps from a tumor patient using 5 UTE-MRF reps and a constant TE of 0.05 ms (resolution = 0.75 x 0.75 mm$^2$). Soft tissues in bone enhanced images are highly suppressed compared to the longitudinal magnetization maps in Fig. 8. The structures of skull produced by UTE-MRF are in good agreement with the CT images, especially the bone density variations indicated by the red arrows.

**Supporting information**

Supporting Information Figure S1. Spectrums of water and oil tubes. The first row (water and oil tubes) shows two major peaks, one is from water and the other is from oil. Only oil sample is scanned in the second row, where multiple peaks could be observed.

Supporting Information Figure S2. Brain quantifications from 3 orthogonal planes in one healthy volunteer using the same in vivo imaging parameters as the main text. The sponges around the ears and below the head are detected and appear dark in $T_1$ maps and bright in bone enhanced images. Although TE is minimized to 50 μs to maximize SNR, UTE-MRF loses its sensitivity to short and ultrashort $T_2$ tissues (Fig. 2). Therefore, the $T_2$ of sponges and bone is misestimated. There are large $T_2$ errors at nasal cavity, ventricle, and skull basis especially in sagittal scan. This is highly probably caused by the flow induced signal mismatching in ventricle and the absence of signal in nasal cavity and skull bases.

Supporting Information TABLE S1. The Comparisons between the Nominal Relaxation Times and the Average Measurements from UTE-MRF.

Supporting Information Video S1. The comparisons between the phase images (under 6 times of repetitions) before (left) and after (right) sliding window (window size = 20). Strong radial undersampling streaking artifact at left phase images is substantially reduced after sliding window. Therefore, radial undersampling aliasing contributes to most of the noise term at equation [6] at the main text. Thus the noise term in phase images is ignored after sliding window.

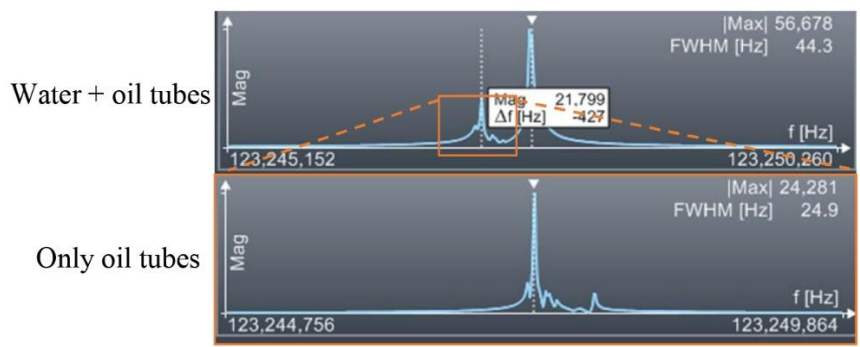

Supporting Information Figure S1. Spectrums of water and oil tubes. The first row (water and oil tubes) shows two major peaks, one is from water and the other is from oil. Only oil sample is scanned in the second row, where multiple peaks could be observed.

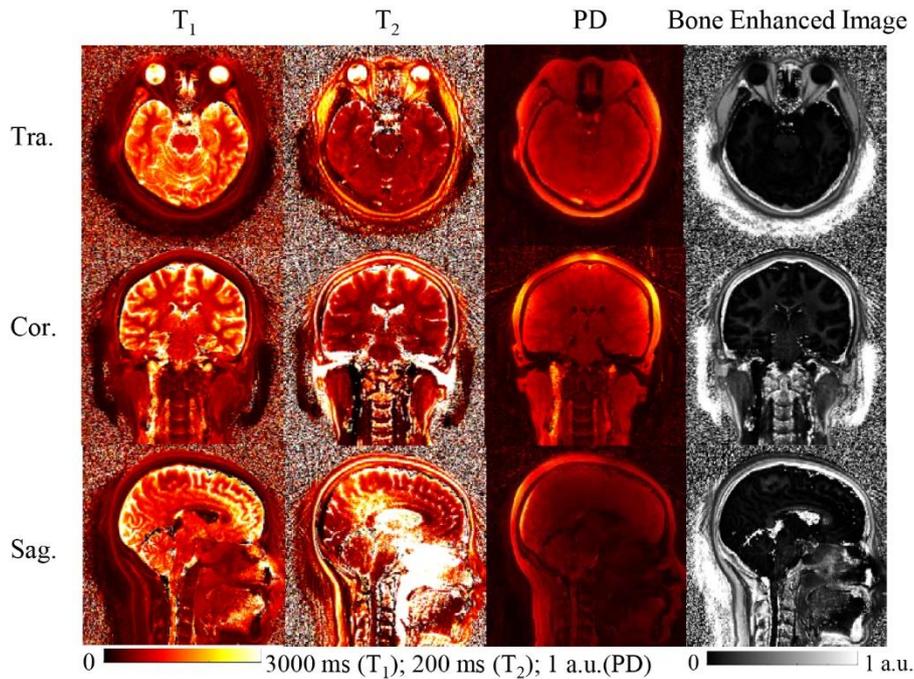

Supporting Information Figure S2. Brain quantifications from 3 orthogonal planes in one healthy volunteer using the same in vivo imaging parameters as the main text. The sponges around the ears and below the head are detected and appear dark in $T_1$ maps and bright in bone enhanced images. Although TE is minimized to 50 µs to maximize SNR, UTE-MRF loses its sensitivity to short and ultrashort $T_2$ tissues (Fig. 2). Therefore, the $T_2$ of sponges and bone is misestimated. There are large $T_2$ errors at nasal cavity, ventricle, and skull basis especially in sagittal scan. This is highly probably caused by the flow induced signal mismatching in ventricle and the absence of signal in nasal cavity and skull bases.

Supporting Information TABLE S1. The Comparisons between the Nominal Relaxation Times and the Average Measurements from UTE-MRF

| (ms) | Muscle | Tendon | Total water | Free water |
|---|---|---|---|---|
| **Nominal $T_1$** | 1400 | 621 | 246 | 524 |
| **UTE-MRF $T_1$** | 1395 | 619 | 245 | 523 |
| **Nominal $T_2/T_2'/T_2^*$** | 32/114/25 | 3/9.9/2.3 | 1.2/1.7/0.7 | 3.5/7.6/2.4 |
| **UTE-MRF $T_2$** | 32.3 | 2.6 | 0.7 | 3.4 |